\documentclass[a4paper,11pt]{article}
\pdfoutput=1 
\usepackage{jcappub,xcolor} 

\usepackage{amssymb,amsmath}
\usepackage{bm}

\usepackage{footnote}

\newcommand{\beq}{\begin{equation}}
\newcommand{\beqn}{\begin{eqnarray}}
\newcommand{\eeq}{\end{equation}}
\newcommand{\eeqn}{\end{eqnarray}}

\newcommand{\ud}{\,\mathrm{d}}

\newcommand{\atanh}{\mathop{\mathrm{tanh^{-1}}}}

\usepackage{verbatim}

\begin{document}

\begin{flushright}
Nikhef 2018-014
\end{flushright}

\title{\center \rm \bf \Huge Angular inflation \\ in multi-field $\alpha$-attractors}

\author[a]{Perseas Christodoulidis,}
\author[a]{Diederik Roest,}
\author[b,c]{Evangelos I. Sfakianakis}
\affiliation[a]{Van Swinderen Institute for Particle Physics and Gravity, 
University of Groningen, Nijenborgh 4, 9747 AG Groningen, The Netherlands}
\affiliation[b]{Nikhef, Science Park 105, 1098 XG Amsterdam, The Netherlands}
\affiliation[c]{Lorentz Institute for theoretical physics, University of Leiden, 2333CA Leiden, The Netherlands}
\emailAdd{p.christodoulidis@rug.nl; d.roest@rug.nl; e.sfakianakis@nikhef.nl}


\abstract{We explore the dynamics of multi-field models of inflation in which the field-space metric is a hyperbolic manifold of constant curvature. Such models are known as $\alpha$-attractors and their single-field regimes have been extensively studied in the context of inflation and supergravity. We find a variety of multi-field inflationary trajectories in different regions of parameter space, which is spanned by the mass parameters and the hyperbolic curvature. Amongst these is a novel dynamical attractor along the boundary of the Poincare disc which we dub ``angular inflation''. We calculate the evolution of adiabatic and isocurvature fluctuations during this regime and show that, while isocurvature modes decay during this phase, the duration of the angular inflation period can shift the single-field predictions of $\alpha$-attractors.
For highly curved field-space manifolds, this can lead to predictions that lie outside the current observational bounds.
}

\maketitle

\section{Introduction} 
In contrast to what the Big Bang theory would have one believe, the Universe is remarkably featureless on scales beyond those of superclusters: cosmic microwave background (CMB) observations as well as large-scale surveys paint a picture of striking isotropy and homogeneity. On top of this smoothness, small fluctuations have given rise to structure formation. These can be observed in their infancy as CMB anisotropies: first seen by the COBE satellite, recent results from the {\it Planck} satellite have provided the most accurate measurements of these fluctuations to date \cite{Ade:2015lrj}.

Currently, our best explanation for this dichotomous beginning of the Universe is provided by the inflationary paradigm \cite{Guth:1980zm,Linde:1981mu}. Its rapid exponential expansion has diluted any primordial stark features, while microscopic quantum fluctuations have seeded the subsequent structure formation. The CMB therefore provides a window on these quantum fluctuations. The aforementioned satellite missions have established the amplitudes of scalar fluctuations, as well as their scale dependence encoded in the spectral index, with $n_s = 0.968 \pm 0.006$. Moreover, there are increasingly strong bounds on the amplitude of tensor fluctuations, captured in the tensor-to-scalar ratio, with $r < 0.07$ \cite{Ade:2015lrj,Ade:2015tva}.

It falls to the theoretical community to model and explain these numbers. While many models with various scalar potentials have been constructed to date, following the Planck 2013 release a different argumentation has been put forward. This employs the non-trivial structure of multi-field kinetic terms. At the two-derivative level, these can be interpreted as a scalar geometry ${\cal G}_{IJ} (\phi)$. Inflation on a curved scalar manifold (see e.g. Refs.~\cite{Peterson:2011yt,Peterson:2010np,Peterson:2010mv}) displays a variety of novel signatures due to geometric effects, including imprints from heavy fields during turns in field space \cite{heavy1, heavy2}, curvature fluctuations from ultra-light entropy modes \cite{Achucarro:2016fby},  as well as inflationary destabilization due to curvature \cite{Renaux-Petel:2015mga}.

Remarkably, in the specific (and highly symmetric) case of a  hyperbolic scalar geometry, one naturally satisfies the Planck bounds on the spectral index and tensor-to-scalar ratio. As a result of the kinetic interactions on the hyperbolic manifold, there is a significant insensitivity to the potential interactions, leading to robust predictions that take the form \cite{Kallosh:2013yoa, Ferrara:2013rsa,Kallosh:2013maa,Kallosh:2013daa,Kallosh:2015lwa,Carrasco:2015rva}
 \begin{align}
   n_s = 1 - \frac2N \,, \qquad r = \frac{12 \alpha}{N^2} \,, \label{universal}
 \end{align}
to leading order in an expansion in the inverse number of e-folds $1/N$, where we consider the CMB-relevant perturbations to have exited the horizon $N$ $e$-folds before the end of inflation. These depend on a single parameter $\alpha$ that sets the hyperbolic curvature. The same predictions can be reached from different perspectives, including Starobinsky's model with $\alpha=1$ \cite{Staro} and non-minimal couplings with $\alpha = 1 + 1/(6\xi)$ \cite{nonm1, KS}. At some level, the unifying feature of all these approaches can be attributed to a singularity in the kinetic sector, whose leading Laurent expansion determines the inflationary predictions \cite{Galante:2014ifa}.

Importantly, the above predictions were derived under the assumption of an effectively single-field trajectory. This can be achieved by the inclusion of higher-order terms that render the orthogonal directions heavy, see e.g. Ref.~\cite{Carrasco}. However, it would be interesting to see what genuine multi-field effects can arise in a more general situation. From a theoretical  perspective, our current understanding of high-energy theories suggests a multitude of scalar fields, without an a priori reason that only one of these should be light. Similarly, from an observational perspective,  such multi-field effects might allow for novel signatures, as has been studied in great detail in a wide variety of models, including N-flation \cite{Dimopoulos:2005ac}, many-field models \cite{Easther:2013rva,Dias:2016slx} or inspired by random matrix theories \cite{Easther:2005zr, Dias:2017gva}.

Very recently, a similar study of multi-field effects in a hyperbolic manifold, as suggested by the {\it Planck} results, was undertaken \cite{Achucarro:2017ing}. Remarkably, it was found that under a set of conditions, even multi-field inflationary trajectories on a hyperbolic manifold adhere to the above universal predictions. For instance, one can consider a scalar potential with a rotationally symmetric mass term and a symmetry-breaking quartic term, both of which live on the Poincare disk (and admit a simple supergravity embedding when $\alpha=1/3$). The background dynamics was found to be almost perfectly radial in a range of parameter values, ``rolling on a ridge'', despite the presence of an angular slope. Moreover, perturbations around this non-trivial background have a remarkable structure that results in predictions identical to those in Eq.~\eqref{universal} despite the presence of multi-field effects.

This paper will build on previous work by investigating the multi-field behaviour of $\alpha$-attractors in a wide range of parameter space, and by pointing out multi-field effects that go beyond the universal behaviour. As the simplest possible case, we will study quadratic potentials on the Poincare disk and consider a range of masses and hyperbolic curvatures. In addition to the above radial dynamics with universal behaviour, we will display a second regime of inflation that proceeds along the angular direction. Instead of the radial dynamics, which correspond to the slow-roll approximation in all coordinates, angular inflation only employs slow-roll for the angular coordinate, while the radial coordinate to a first approximation is frozen. We will outline when this novel regime appears, why it can be thought of as an alternative attractor\footnote{One might think that this regime is related to that of hyperinflation \cite{Brown:2017osf, Mizuno:2017idt}, which also crucially relies on the hyperbolic manifold. However, hyperinflation is an alternative to slow-roll inflation in the case of a very steep and rotationally symmetric potential.}, and how it modifies the duration and predictions of inflation\footnote{We have been informed about a forthcoming publication with possibly related effects \cite{Linde:2018hmx}.}.

This paper is organized as follows. In Section~\ref{section:background} we describe the model and compute the background evolution for a wide range of parameter combinations, discovering qualitatively different behaviour in different parts of parameter space. The perturbations are discussed in Section~\ref{sec:perturbations}, where we display the formalism and compute all relevant quantities, both analytically and numerically, that define the super-horizon evolution of the adiabatic and isocurvature modes. 
The connection to the observational data is given in Section~\ref{subsec:observables}, where we show that a prolonged period of angular inflation can significantly modify the spectral observables given in Eq.~\eqref{universal} or even place them outside the current bounds from {\it Planck}. However, the change is simple to understand intuitively and to compute analytically, since it only hinges on the number of $e$-folds of non-radial inflation. 
We conclude in Section~\ref{sec:summary}.

\noindent {\bf Conventions:} 
Greek letters label spacetime indices, $\mu,\nu = 0, 1, 2, 3$, with spacetime metric signature $(-,+++)$. Lower-case latin letters label space indices $i,j=1,2,3$ and upper-case latin letters label field-space indices, $I,J =1,2$. We work in terms of the reduced Planck mass, $M_{\rm Pl} \equiv (8\pi G)^{-1/2} = 2.43\times 10^{18} \, {\rm GeV}$.

\section{Background evolution}
\label{section:background}

\subsection{The multi-field $\alpha$-attractor model}
We consider a number of scalar fields that are minimally coupled to gravity, with
\beq
S = \int d^4x \sqrt{-g} \left [ {M_{\rm Pl}^2\over2} R - {1\over 2}{\cal G}_{IJ} g^{\mu\nu} \partial_\mu \phi^I \partial_\nu \phi^J - V(\phi^I)\right ] \,.
\label{eq:action}
\eeq
The $00$ and $0i$ components of the Einstein equations at background order are
\beqn
&& H^2 = {1\over 3M_{\rm Pl}^2 }\left [ {1\over 2}{\cal G}_{IJ} \dot \phi^I \dot \phi^J +V(\phi^I)
\right] \,, \qquad
\dot H = -{1\over 2M_{\rm Pl}^2} {\cal G}_{IJ} \dot \phi^I \dot \phi^J \,.
\eeqn
and the equations of motion for the fields are 
\beq \label{eom}
{\cal D}_t\dot \varphi^I + 3H\dot \varphi^I + {\cal G}^{IJ} V_{,J}=0 \, ,
\eeq
where ${\cal D}_t A^I \equiv \dot \varphi^J {\cal D}_J A^I$ is the directional covariant derivative. The case of a large number of fields is especially interesting in  view of high energy theories, such as string theory, resulting in a large number of scalar fields at high energies. However, in order to make the system tractable and be able to extract all its features in the various parameter regimes, we limit ourselves to only two fields. We leave the many-field generalization for future work.

The two fields $\phi^1\equiv \phi$ and $\phi^2\equiv \chi$ take values on the Poincare disk with the field-space metric
\beq
{\cal G}_{IJ} ={6 \alpha \over (1-\phi^2 -\chi^2)^2} \delta_{IJ} \, ,
\eeq
in which the fields are dimensionless, since we are measuring both the fields and the curvature parameter $\alpha$ in units of $M_{\rm Pl}$.
 We use a quadratic potential,  due to its simplicity and as a first step towards a generalization of extensive studies of multi-field effects in inflation  \cite{Frazer:2013zoa}:
\beq
\label{eq:potential}
V(\phi^I) = {\alpha\over 2} \left (
m_\phi^2 \phi^2 +m_\chi^2 \chi^2 \right ) \,, \quad R_{\rm m}\equiv m_\chi ^2/ m_\phi ^2 \,.
\eeq
For concreteness we will always consider $R_{\rm m}>1$. Note that the factor $\alpha$ has been inserted to give the potential the correct dimensions and to recover the usual quadratic potential in the flat limit $\alpha \rightarrow \infty$ (after the appropriate rescaling of the fields). 
 
Due to the spherical symmetry of the field-space manifold, it will be useful to introduce polar coordinates $\phi = r \cos(\theta)$ and $\chi = r \sin(\theta)$.
In this parametrization of the hyperbolic geometry, the scalar field equations read
\beqn
\ddot r  +{ 2 r \dot r^2 \over 1-r^2} -{ r( r^2 + 1) \over 1-r^2}\dot \theta^2  + 3 H \dot r  + {1  \over 6}\left( m_\phi^2 \cos^2\theta + m_\chi^2 \sin^2\theta  \right)(1-r^2)^2 r &=&0  \, ,
\label{eq:r}
\\
\ddot \theta + {2 (1+r^2) \over r (1-r^2) } \dot r \dot\theta  +
3H\dot \theta +\frac{1}{12} (m_\chi^2-m_\phi^2) \left(1-r^2\right)^2 \sin (2 \theta ) &=&0 \,,
\label{eq:theta}
\eeqn
where we have assumed the spatial homogeneity of the FRW background. In some of the following plots we will use the canonical radius $\psi=\sqrt{6 \alpha}\atanh(r)$ to visualize the different regimes of the fields' evolution, with horizontal and vertical components $\psi \cos(\theta)$ and $\psi \sin(\theta)$.

\subsection{Radial slow-roll inflation}

Starting with the radial equation of motion, it was shown in Ref.~\cite{Achucarro:2017ing} that for initial conditions placing the two fields close to the boundary of the Poincare disc, a period of radial inflation with $\theta(t) \simeq \theta_0$ is supported, where $\theta_0$ is the initial value of the angle in field-space. Provided the initial velocities are sufficiently small, a phase of slow-roll inflation ensues during which the gradient term is counterbalanced by the Hubble friction. Intuitively, this can be understood as the Christoffel terms depend quadratically on the velocities whereas the accelerations can be viewed as finite differences of small quantities.

\begin{figure}[t!]
\includegraphics[width=0.49\textwidth]{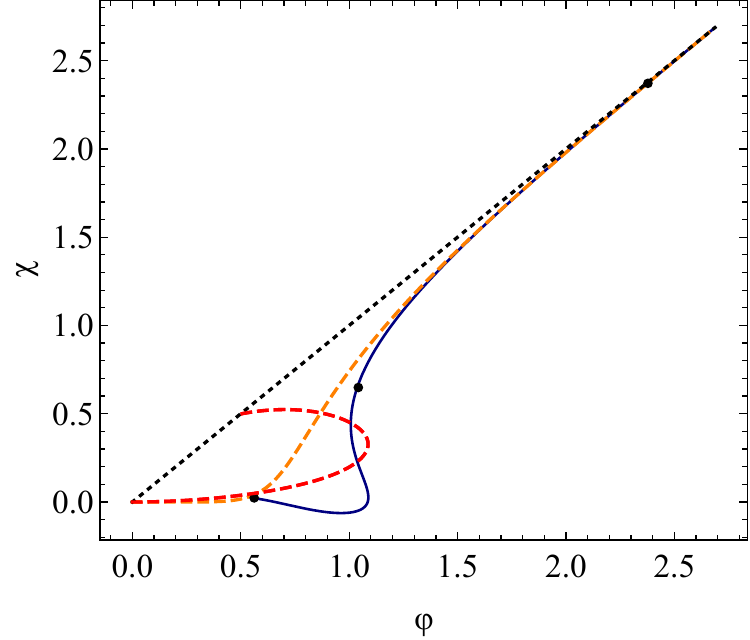} 
\includegraphics[width=0.5\textwidth]{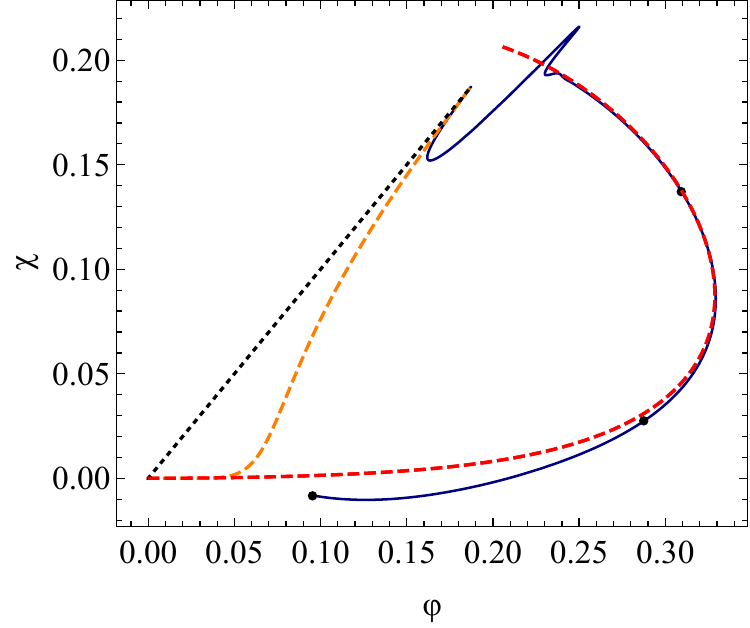} 
\caption{ \it
The evolution of the system in terms of the canonically normalized radius $\psi$ using the full equations of motion (blue) and using the analytic slow-roll expression (orange-dashed) for the parameter values $R_{\rm m} =9$, $\theta_0=\pi/4$, and $\alpha=1/6$, $r_0=0.999$ (left panel) and $\alpha =1/600$, $r_0=0.99$ (right panel). The red dashed line in the right panel is the angular inflation approximation. The black dotted line is the diagonal and the black dots correspond to 55 (CMB point), 2 and zero e-folds before the end of inflation.}
\label{fig:radial_angular_example}
\end{figure}

More precisely, the range of validity of the slow-roll approximation can be measured by the smallness of the multifield generalization of the single field slow roll parameters \cite{Lyth:2009zz,Yang:2012bs}
\begin{align}\label{multiepsilon}
\epsilon_V &= \frac{1}{2V^2} G^{IJ} V_{,I} V_{,J} \,, \qquad
\eta_V = \frac{||V^{,K}\nabla_K V^{,I}||}{V||V^{,I}||}.
\end{align}
If both parameters remain small up to the last few efolds, then the evolution equations can be approximated by slow roll\footnote{The definition of $\epsilon$ given in Eq.~\eqref{multiepsilon} is valid for the slow-roll-slow-turn regime. In general one must use $\epsilon = -\dot H /H^2$, which is what we do for the remainder of this work. For a thorough discussion of the slow-roll and slow-turn approximations see e.g. Ref.~\cite{Peterson:2011yt,Peterson:2010np,Peterson:2010mv}.} and the end of inflation can be estimated by $\epsilon_V \approx 1$. Using the slow roll expressions for the velocities
\begin{equation}
\frac{\dot{r}_{\rm SR}}{\dot{\theta}_{\rm SR}} = \frac{V^{,r}}{V^{,\theta}} \, ,
\end{equation}
we can obtain the relation $r(\theta)$. Since the metric is conformally flat it drops out from the above expression and the parametric relation $r(\theta)$ has exactly the same form as in the flat case. It is straightforward to calculate the number of efolds\footnote{Note that our model, with a  product-separable potential (or sum separable when written in Cartesian coordinates) plus a conformally flat metric, is a third example of models which admits analytic calculation of the number of efolds in the slow-roll slow-turn approximation, along with the sum- or product-separable potentials in flat space \cite{Starobinsky:1986fxa,GarciaBellido:1995qq}. }:
\begin{equation} \label{nefsr}
N_{\rm radial} = \int \ud r \frac{V}{V^{,r}} = \frac{3\alpha}{2} \left( \frac{1}{1-r^2} - \frac{1}{1-r_{\rm end}^2} \right).
\end{equation}
The number of radial e-folds has the same form as in single-field $\alpha$-attractors \cite{Kallosh:2013yoa} and is controlled by the curvature of field-space and the proximity of the initial conditions to the boundary of the Poincare disc. At the end of inflation, given the fact that the potential has a light direction, the system will have to relax in such a way, as to evolve along the light direction. For the parameter range considered in Ref.~\cite{Achucarro:2017ing}, where both the angular gradients and the field space curvature were small (this statement will be quantified shortly), this relaxation towards the light field direction only lasts for one or two $e$-folds. An illustration of this can be found in the left panel of Fig.~\ref{fig:radial_angular_example}. Before we proceed we must note that, although the trajectories considered in Ref.~\cite{Achucarro:2017ing} are highly radial, the existence of a non-zero turn rate is important and one cannot neglect multi-field effects.

However, the validity of the slow-roll approximation breaks down earlier as one considers large hyperbolic curvatures. The first slow-roll parameter is given by
\begin{equation*}
\epsilon_{\rm curved} = \frac{\left( 1 - r^2\right)^2 }{6 \alpha} \epsilon_{\rm flat} ,
\end{equation*}
where $\epsilon_{\rm flat}$ is greater than $1$ (inside the Poincare disc) and diverges when $r\rightarrow 0$.  For small $\alpha $  the condition $\epsilon_{\rm curved}  <1 $ implies that the fields must inflate close to the boundary and at the end of inflation $ 1 - r_{\rm end}^2 \sim \sqrt{\alpha}$ should hold.  At the same time $\eta$, which is given by a much more complicated expression, contains terms that scale as $\left( 1 - r_{\rm end}^2\right)/ \alpha$. By comparing the two we conclude that when $\alpha$ is small, there will be a region before the end of inflation in which slow-roll fails. This is the regime that we will turn to next.

\subsection{Angular inflation}

It follows from the scalar field equations \eqref{eq:r} and \eqref{eq:theta} that the only terms which are enhanced by the field-space curvature close to the boundary of the Poincare disc are the Christoffel terms. However, they are velocity-suppressed   during radial slow-roll  inflation. Towards the end of radial inflation, the fields speed up and the increase in $\dot r$ boosts the Christoffel terms (leading to related trajectories in e.g. Refs.~\cite{Bond:2006nc, Kallosh:2014qta,Lazaroiu:2017avq}). 

This gives rise to two competing effects; the geodesic motion aims to push both fields to the boundary of the disk following a circular arc, whereas the gradient of the potential attracts both fields to the origin. Unlike $\dot{\theta}$, which becomes zero close to the minimum of the potential at $\theta=0,\pi/2$, $\dot{r}$ can vanish away from the minimum because of the presence of $\Gamma^{r}_{\phi \phi}$. A solution with $\dot{r}\approx 0$ can be sustained either by sufficiently reducing  $\alpha$ (so that inflation takes place closer to the boundary of the space where the Christoffel $\Gamma^r_{\theta \theta}$ becomes important) or by increasing the mass ratio (and hence the velocity $\dot{\theta}^2$). An illustration of this can be found in the right panel of Fig.~\ref{fig:radial_angular_example}, which shows the non-monotonic behaviour of the radial coordinate. Moreover, it highlights that the trajectory proceeds for a significant portion along an angular direction, i.e.~with $\dot r$ nearly vanishing.

\begin{figure}[t]
\includegraphics[width=0.5\textwidth]{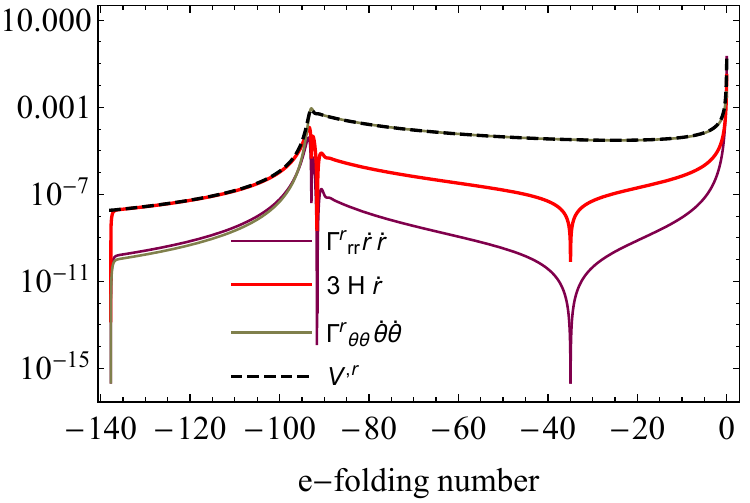} 
\includegraphics[width=0.5\textwidth]{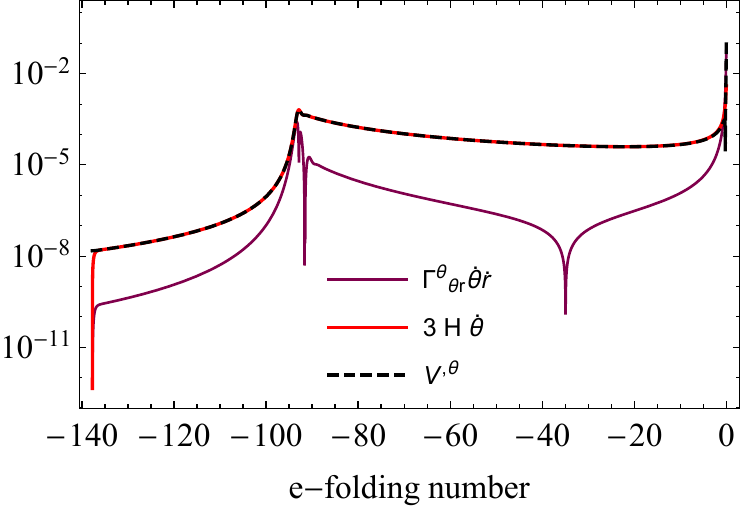} 
\caption{\it Numerical comparison of various terms in equations \eqref{eq:r} (left) and \eqref{eq:theta} (right) for $\alpha=1/600$, $R_{\rm m}=9$, $\theta_0=\pi/4$ and $r_0=0.99997$. We have chosen these conditions to support both a period of radial as well as a angular inflation. The dominant terms in each period are indeed those corresponding to the relevant approximations.}
\label{fig:dominantterms}
\end{figure}

This can be understood by considering the third and last term of Eq.~\eqref{eq:r}, which do not depend on $\dot{r}$ and  have opposite signs. If they almost cancel each other then the radius will be a slowly varying function and therefore fields will perform a predominately angular motion. Indeed, one can show numerically that the dominant terms of the radial and angular equations of  motion are
\beqn \label{eq:ang_theta}
\frac{\left(2 r^3+2 r\right)}{2 r^2-2} \dot \theta^2+ {1\over 12} \left [ m_\chi^2 + m_\phi^2 + (m_\phi^2 -m_\chi^2) \cos(2\theta) \right ] (1-r^2)^2 r &\approx&0 \, ,
\\\label{eq:ang_r}
3H\dot \theta +\frac{1}{12} (m_\chi^2-m_\phi^2) \left(1-r^2\right)^2 \sin (2 \theta ) &\approx&0\, .
\eeqn
Hence the angular motion is dominated by the usual slow-roll combination of the potential and Hubble drag terms, while the radial equation instead is dominated by the potential and the $\Gamma_{\theta\theta}^r$ Christoffel term, with all other terms being subdominant. Fig.~\ref{fig:dominantterms} shows the  magnitude of the various terms for a characteristic choice of parameters, supporting a prolonged period of angular motion.
We see that our approximations leading to Eqs.~\eqref{eq:ang_theta} and \eqref{eq:ang_r} are justified.

\begin{figure}[t]
\centering
\includegraphics[width=0.45\textwidth]{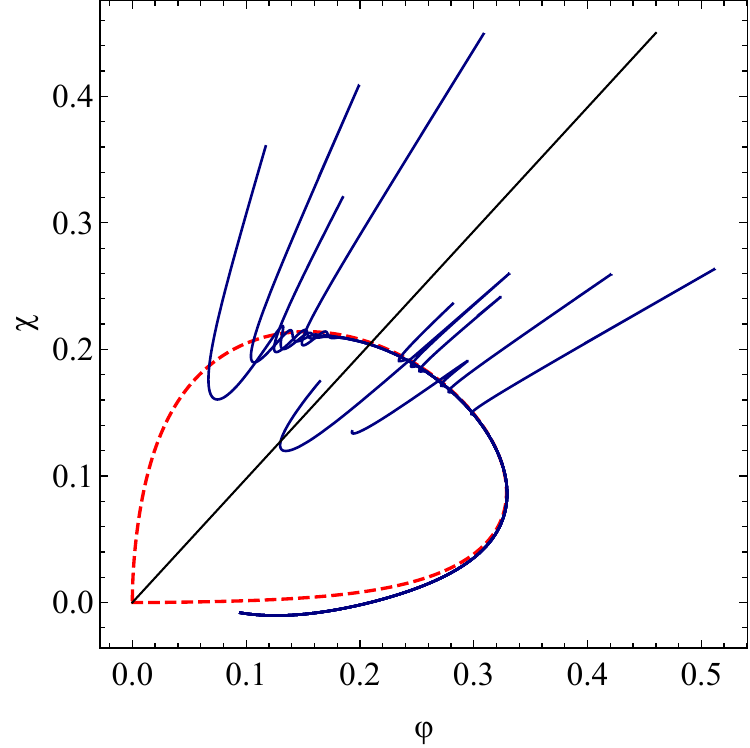} 
\caption{\it
Plot of the angular attractor for $\alpha=1/600$, $R_{\rm m}=9$ and a wide range of initial data (the black line is the diagonal). We observe that the angular solution captures very well the evolution for $\theta<\pi/4$.
}
\label{fig:ang_attractor}
\end{figure}

The regime of angular inflation allows for a simple derivation of the trajectory through field space. One can eliminate  $\dot \theta$ from the approximate equations \eqref{eq:ang_theta} and \eqref{eq:ang_r}. Under the additional assumption that the Hubble function is dominated by the potential $H^2 \simeq \tfrac13 V$ (the neglect of the kinetic terms can easily be justified as the angular motion is slow-roll, while the radial motion is even slower), one obtains a parametric equation for the field trajectory during angular motion
\beq
r(\theta)=\frac{ \sqrt{ \sqrt{\frac{81}{4} \alpha ^2 ((1-R_{\rm m}) \cos (2 \theta )+R_{\rm m}+1)^4+ (1-v)^4 \sin ^4(2 \theta )}-{9\over 2} \alpha  ((1-R_{\rm m}) \cos (2 \theta )+ R_{\rm m}+1)^2}}{ \sin (2 \theta ) (R_{\rm m}-1)} \, ,
\label{eq:rthetafull}
\eeq
where $R_{\rm m}$ is defined in Eq.~\eqref{eq:potential}. We have verified the attracting nature of this solution for a wide range of initial conditions in Fig.~\ref{fig:ang_attractor}. 
The remarkable property of this solution is that it only depends on $\alpha$, the mass ratio $R_{\rm m}$ and the initial angle $\theta_0$, while all other initial data have dropped out. In contrast, for the slow roll approximation (or any other approximation that only fixes the velocities), one obtains a family of trajectories through multi-dimensional field space - which is e.g.~a one-parameter family for the two-field case. In a sense, the angular inflation solution is therefore closer to the notion of single-field attractor as it is independent of all initial conditions \footnote{This type of coordinate dependent approximation, like the case of hyperinflation, is more obscure in Cartesian coordinates because the Christoffel terms of the light field are not balanced by the gradient term. The attractor solution is, of course, independent of coordinate system but an approximate form would be very hard to find.}.

While it captures the angular inflation regime very well and has the remarkable propery of a dynamical attractor, the solution $r(\theta)$ it is somewhat complicated and difficult to use. Later on we will often use the quantity $1-r^2$, which encodes the stretching of the field-space as one nears the boundary of the Poincare disk. Expanding in terms of $\alpha$, this can be written as
\beq
1-r^2 \simeq \frac{9 \alpha  (\cot \theta + R_{\rm m}\tan \theta  )^2}{2 (R_{\rm m}-1)^2} \,.
\label{eq:rthetaapprox2}
\eeq 
This relation breaks down close to the two axes. It is easy to see that unless the initial angle is close to the heavy field direction $\theta \approx \pi/2$, the two expressions given by Eqs.~\eqref{eq:rthetafull} and \eqref{eq:rthetaapprox2} match very well. There is some disagreement close to the light direction $\theta=0$, but there we expect the slow-roll approximation to break either way, so the comparison is meaningless. From now on, we will always use the approximation of Eq.~\eqref{eq:rthetaapprox2}, unless otherwise noted.

We now move to computing the first slow-roll parameter $\epsilon$. Using the above approximations, amounting to $\dot{r}\approx 0$, $H^2 \approx {1\over 3} V$ and $\dot{\theta}=\dot{\theta}_{\rm SR}$, one has
\beq
\epsilon = -{ \dot H \over H^2} \approx {3 \over 2} (1-r^2)\, ,
\label{eq:epsilontheta}
\eeq 
where we expanded this relation close to the boundary $r=1$. The minimum value of $\epsilon$ during the angular motion (in the small $\alpha$ approximation) occurs at and is given by
\beq
\theta_{\epsilon,{\rm min}} = \arctan \left({1\over \sqrt{R_{\rm m}}} \right) \,, \qquad \epsilon_{\rm min} = \frac{27 \alpha R_{\rm m}}{ (R_{\rm m}-1)^2} \,.
\label{eq:epsilonmin}
\eeq
This is significantly higher than the value of $\epsilon$ during inflation along the radial direction\footnote{The different value of $\epsilon$ has a significant effect for modes that are exiting the horizon during the angular part of inflation. The analysis of these modes will presented in a separate publication.}.

Using the expression for $\epsilon$ we can compute the angle at which angular inflation ends.
The equation $\epsilon=1$ can be solved, using Eq.~\eqref{eq:epsilontheta}, however the solution is not very insightful. At small $\alpha$ and large $R$ it reads
\beq
\theta_{\rm end} = \frac{3 \sqrt{3 \alpha }}{2 (R_{\rm m}-1)}\, .
\eeq
which is indeed close to $\theta=0$. Since close to $\epsilon=1$ our approximations break down, for small $\alpha$ and / or large mass ratio we can safely take angular inflation to end for $\theta=0$, without introducing extra errors\footnote{In fact, computing $N(\theta=\theta_{\rm end})$ and $N(\theta=0)$ gives agreement to within a few decimal points, so we will  $\theta=0$ as the end-point of angular inflation.}.

\begin{figure}[t]
\centering
\includegraphics[width=0.45\textwidth]{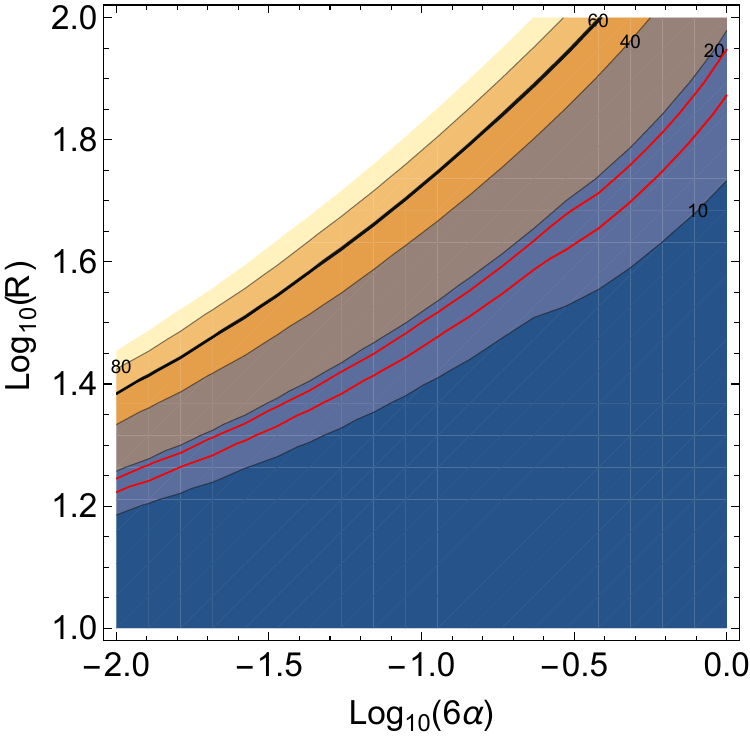}
\includegraphics[width=0.45\textwidth]{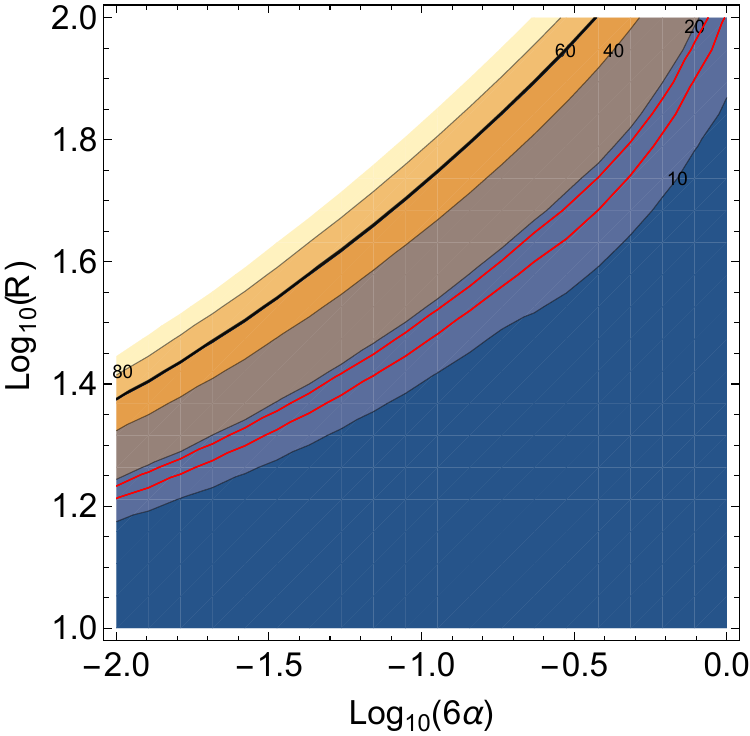} 
\caption{\it 
The number of $e$-folds, occurring after the initial period of radial inflation for a wide range of the field-space curvature, characterized by $\alpha$ and the mass ratio of the two fields $R_{\rm m}$ for initial angles $\theta_0=\pi/4$ (left) and $\theta_0=\pi/3$ (right). The thick black line corresponds to 60 e-folds of non-radial inflation and the red to the Planck contours to the point where the predictions of our model exit the Planck $1\sigma$ and $2\sigma$ regions, as computed in Eq.~\eqref{offset} and discussed in Section \ref{sec:perturbations}.
The two panels are very similar since small changes in the initial angle do not alter the duration of the angular regime significantly. This can be seen in the expression for the number of $e$-folds of angular inflation in Eq.~\eqref{eq:seriesN}, where the $\theta$-dependence is not present in the  lowest order term.
}
\label{fig:Ncontourplot}
\end{figure}

Finally, computing the number of e-folds during angular motion is straightforward using the angular slow-roll approximations for $H$ and $\dot\theta$
\beq
N =\int_{t_0}^{t_{\rm end}} H dt = \int_{\theta_0}^{\theta_{\rm end}} {H\over \dot \theta} d\theta\, .
\eeq
The integration is performed from the initial angle $\theta$ until the final angle which we take to be $\theta=0$. The result is
\beq
N=-\frac{\frac{8 \left(-\left(R_{\rm m}^2-1\right) \cos (2 \theta )+R_{\rm m}^2+R_{\rm m}+1\right)}{(-(R_{\rm m}-1) \cos (2 \theta )+R_{\rm m}+1)^2}+(18 \alpha +4) \log \left(\frac{1}{2} (-(R_{\rm m}-1) \cos (2 \theta )+R_{\rm m}+1)\right)-2 (R_{\rm m}+2)}{54 \alpha }\, ,
\label{eq:Nangular}
\eeq
This relation outlines where in parameter space there is a significant number of e-folds during angular inflation. Fig.~\ref{fig:Ncontourplot} shows the number of non-radial e-folds for a broad range of parameters. We see that for large $R_{\rm m}/\alpha$, the contours of $N$ give a linear relation between $\log(R_{\rm m})$ and $\log(\alpha)$. This can also be calculated by expanding the above in $\alpha\ll1$ and $R_{\rm m} \gg 1$:
\beqn
\label{eq:seriesN}
N&\approx &{ R_{\rm m}\over 27 \alpha}
+
\frac{-4 \csc ^2\left(\theta \right)-(18 \alpha +4) \log \left(R_{\rm m} \sin ^2\left(\theta \right)\right)+4}{54 \alpha } + \ldots \, .
\eeqn

While it is non-trivial to invert the number of e-folds and get the function $\theta(N)$, some analytical progress can be made towards this goal.
Specfically, we can neglect the logarithms in the function of $N(\theta)$ after which we obtain 
\beq
\theta(N) \simeq \frac{1}{2} \cos ^{-1}\left(\frac{2 R_{\rm m}^2+ R_{\rm m} (2-54 \alpha  N)-2 \sqrt{108 \alpha  R_{\rm m} N+4}- 54 \alpha  N}{(R_{\rm m}-1) (2 R_{\rm m}-54 \alpha  N+4)}\right)\, ,
\eeq
where we took $\theta_0=\pi/4$. One can check that this an increasingly good approximation for larger $R_{\rm m}$.

\begin{figure}[t]
\centering
\includegraphics[width=0.5\textwidth]{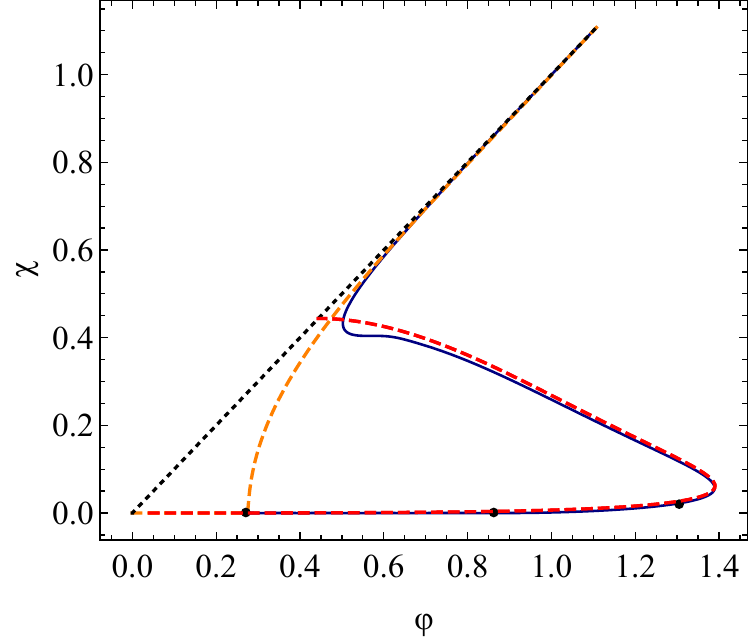} 
\caption{
\it The evolution of the system in terms of the canonically normalized field $\psi$ with identical colour-coding and parameters as in Fig.~\ref{fig:radial_angular_example} but with $R_{\rm m}=500$, $\alpha=1/60$, $\theta_0=\pi/4$ and $r_0=0.9999$. Note that this plot contains all possible different kinds of evolution and at the time of generation of fluctuations  the evolution of the system can be well described by a single field.
}
\label{fig:example2}
\end{figure}

It is important to note that the number of e-folds quoted above also includes the final stage of inflation, which is more like single-field inflation along the horizontal axis (corresponding to the lightest field) than angular inflation along the boundary. This is illustrated in Fig.~\ref{fig:example2}. This feature is particularly noticeable in the case where the mass ratio is large: in this situation there is a sizeable number of e-folds produced in this final stage of inflation. In a somewhat rough sense, the two parameters of our model trigger the two regimes: smaller values of $\alpha$ increase the number of e-folds during the angular trajectory, while a larger mass ratio does the same for the single-field regime along the light axis prior to the end of inflation. For very large mass ratios, the number of e-folds during this final stage exceeds that of the CMB window, and as a result the observable part of inflation will be effectively single-field (with a very heavy orthogonal direction). In this case the inflationary predictions will coincide with the single-field ones. In all other cases one needs to evolve the quantum fluctuations during the angular regime, which is the topic that we turn to next. 

\section{Perturbation analysis}
\label{sec:perturbations}

\subsection{Generalities on adiabatic and isocurvature modes}

We will use the notation and formalism described in Ref.~\cite{KMS} and further utilized in Refs.~\cite{GKS, KS, SSK}, which allows for the computation of perturbations produced in multi field inflation with non-trivial field space metric. We present here the main results and formulas needed for our analysis\footnote{Ref.~\cite{Achucarro:2017ing} used somewhat different notation, but the method and results are easily translatable. } and refer the interested reader to Ref.~\cite{KMS} and references therein, such as Refs.~\cite{GongMultifield, Peterson:2011yt,Peterson:2010np,Peterson:2010mv}.

The evolution equation for the perturbations are
\beq
{\cal D}_t^2{ Q}^I
 + 3H{\cal D}_t { Q}^I + 
 \left [  {k^2 \over a^2} \delta^I_{\, J} + {\cal M}^I_{\, J} - {1\over M_{\rm Pl}^2 a^3} {\cal D}_t\left ( {a^3\over H} \dot \phi^I \dot \phi_J \right) \right ]Q^J=0
 \label{eq:QIeom}\, ,
\eeq
where $Q^I$ are the gauge-invariant Mukhanov-Sasaki variables defined as
\beq
Q^I = \delta \phi^I +{ \dot \phi^I \over H}\psi \,,
\eeq
to lowest order in fluctuations. The effective mass matrix $ {\cal M}^I_{\, J} $ is
\beq
\label{eq:MIJ}
 {\cal M}^I_{\, J} \equiv {\cal G}^{IK} {\cal D}_J{\cal D}_K V - {\cal R}^I_{LMJ} \dot \phi^L \dot \phi^M\, ,
\eeq
where ${\cal R}^I_{LMJ}$ is the Riemann tensor of the field-space manifold.

The analysis of the scalar fluctuations will be aided by decomposing the background fields' velocity vector into a magnitude and a directional unit vector 
\begin{equation}
\dot{\sigma} \equiv |\dot{\varphi}^I| = \sqrt{\mathcal{G}_{IJ} \dot{\varphi}^I \dot{\varphi}^J} \,, \qquad
\hat{\sigma}^I \equiv \frac{\dot{\varphi}^I}{\dot{\sigma}} \,.
\end{equation}
An equally important dynamical quantity is the turn-rate of the background fields, given by 
\begin{equation}
\omega^I = \mathcal{D}_t \hat{\sigma}^I ,
\end{equation}
with magnitude $\omega^2 = {\cal G}_{IJ}\omega^I \omega^J$ and directional unit vector 
\beq
\hat{s}^I \equiv{ \omega^I \over \omega}\, .
\eeq
Using the background equations of motion we can express the magnitude of the turn-rate in terms of the potential 
\beq
|\omega|^2 = {\cal G}_{IJ} \omega^I \omega ^J = {{\cal G}^{IJ} V_I V_J  -(V_\sigma)^2\over \dot\sigma^2}\, .
\eeq
As they are orthogonal, $\hat{s}^I \hat{\sigma}_I = 0$, the unit vectors $\hat{\sigma}^I$ and $\hat{s}^I$ effectively act like projection vectors. In this way, we may decompose the vector of fluctuations $Q^I$ into adiabatic modes $Q_{\sigma} \equiv \hat{\sigma}_I Q^I$ and isocurvature ones $Q_s \equiv \hat{s}_I Q^I$.
The two modes are subject to the following equations of motion:
\begin{equation}
\label{adi}
\begin{split}
\ddot{Q}_{\sigma} &+ 3 H \dot{Q}_{\sigma} + \left[ \frac{k^2}{a^2} + \mathcal{M}_{\sigma \sigma} - \omega^2 - \frac{1}{M_{\text{\rm pl}}^2 a^3} \frac{d}{dt} \left(\frac{a^3 \dot{\sigma}^2}{H}\right)\right] Q_{\sigma}\\ 
& = 2\, \frac{d}{dt} \left(\omega\, Q_s\right) - 2 \left( \frac{V_{,\sigma}}{\dot{\sigma}} + \frac{\dot{H}}{H} \right) \omega\, Q_s ,
\end{split}
\end{equation}
\begin{equation}
\label{ent}
\ddot{Q}_s + 3 H \dot{Q}_s + \left[ \frac{k^2}{a^2} + \mathcal{M}_{ss} + 3 \omega^2 \right] Q_s = 4 M_{\text{\rm pl}}^2 \,\frac{\omega}{\dot{\sigma}} \frac{ k^2}{a^2} \Psi,
\end{equation}
where $\Psi$ is the gauge-invariant Bardeen potential \cite{Malik:2008im, Bassett:2005xm},
\begin{equation}
\Psi \equiv \psi + a^2 H \left( \dot{E} - \frac{B}{a}\right) \, .
\end{equation}
Eqs.~\eqref{adi} and \eqref{ent} feature the projections of the mass-squared matrix, $ \mathcal{M}^I_{\> J}$ defined in Eq.~\eqref{eq:MIJ}, along the adiabatic and isocurvature directions, with 
$\mathcal{M}_{\sigma \sigma} = \hat{\sigma}_I \hat{\sigma}^J \mathcal{M}^I_{\> J}$ and $\mathcal{M}_{ss} = \hat{s}_I \hat{s}^J \mathcal{M}^I_{\> J}$ respectively.

It follows from the above discussion that the entropy perturbations will act as a source of adiabatic perturbations, provided that the turn-rate is non-zero. This was discussed in Ref.~\cite{Achucarro:2017ing} for the initial radial inflationary period. We also note that  entropy perturbations with super-horizon wavelength will have an effective mass-squared of 
\begin{equation}
\label{eqn:mass}
\mu^2_s = \mathcal{M}_{ss} + 3 \omega^2.
\end{equation}
Eq.~\eqref{eqn:mass} shows the distinct contributions to the isocurvature effective mass-squared. The contribution coming from the projection of the mass matrix can be positive or negative; note that it includes the curvature term which for a negatively curved manifold can lead to destabilization \cite{Renaux-Petel:2015mga}. Secondly, the contribution from the turn rate is always positive. Hence, in order to achieve a tachyonic growth of isocurvature fluctuations, the turn rate must be small (this statement will be quantified in later parts of this Section).

In the usual fashion \cite{Malik:2008im, Bassett:2005xm}, we may construct the gauge-invariant curvature perturbation, 
\begin{equation}
\mathcal{R}_c \equiv \psi - \frac{H}{(\rho + p)} \delta q\, ,
\end{equation}
where $\rho$ and $p$ are the background-order density and pressure and $\delta q$ is the energy-density flux of the perturbed fluid. In terms of our projected perturbations, we find \cite{KMS} 
\begin{equation}
\mathcal{R}_c =  \frac{H}{\dot{\sigma}} Q_{\sigma} \,, \qquad 
\mathcal{S} \equiv \frac{H}{\dot{\sigma}} Q_s \,,
\end{equation}

where we have moreover defined an analogously normalized isocurvature perturbation.
In the long-wavelength limit, the coupled perturbations obey relations of the form 
\cite{ Malik:2008im, Bassett:2005xm, Gordon:2000hv,GrootNibbelink:2000vx,GrootNibbelink:2001qt,KMS, Wands:2007bd, Wands:2002bn, KaiserTodhunter}: 
\begin{equation}
\begin{aligned}
&\dot{\mathcal{R}_c} \simeq \tilde \alpha H \mathcal{S} \,, \qquad \dot{ \mathcal{S}} \simeq \tilde \beta H \mathcal{S} \,,
\end{aligned}
\end{equation}
which allows us to write the transfer functions as
\begin{equation}
\label{eq:trans}
 T_{\mathcal{RS}} (t_{*}, t) = \int_{t_{*}}^t dt'\,\tilde \alpha(t')\, H(t') \,T_{\mathcal{SS}}(t_{*}, t') \,, \qquad T_{\mathcal{SS}} (t_{*}, t) = \mathrm{exp} \left[ \int_{t_{*}}^t dt'~ \tilde \beta(t')\, H(t') \right],
\end{equation}
where $t_{*}$ is any time after which the mode has left the horizon. Usually it is taken to be
the time when a fiducial scale of interest first crosses the Hubble radius during inflation, $k_{*} = a (t_{*}) H (t_{*})$. However, in the current context, it can be useful to take it to be the time, when the period of initial radial inflation ends.
We find \cite{KMS}
\begin{equation}
\begin{aligned}
&  \tilde \alpha = \frac{2 \omega}{H} \,, \qquad \tilde \beta = -2 \epsilon - \eta_{ss} + \eta_{\sigma \sigma} - \frac{4 \omega^2}{3 H^2},
\end{aligned}
\label{eqn:albet}
\end{equation}
where instead of a single second slow-roll parameter $\eta$ we define the second slow roll (acceleration) parameters along the adiabatic  direction as $\eta_{\sigma \sigma} \equiv {M_{\text{\rm pl}}^2 \,\mathcal{M}_{\sigma \sigma}} / {V}$ and similar for the isocurvature direction\footnote{Although $\eta_{\sigma\sigma}$ and $\eta_{ss}$ are called slow-roll parameters in the literature they are not necessarily small. For the $\sigma$ field a possibly better definition would be $\eta_{\sigma\sigma}=(M_{\sigma\sigma}-\omega^2)/V$ which can be shown to behave as a slow roll parameter when $\ddot{\sigma}$ is small.}.

Using the transfer functions, we may relate the power spectra at some time $t_{*}$ to spectra at later times. In the regime of interest, for late times and long wavelengths, we have
\begin{equation}
\begin{aligned}
&\mathcal{P}_{\mathcal{R}}(k) = \mathcal{P}_{\mathcal{R}}(k_{*}) \left[1 + T_{\mathcal{RS}}^2 (t_{*}, t)\right] \,, \qquad \mathcal{P}_{\mathcal{S}}(k) = \mathcal{P}_{\mathcal{S}}(k_{*})\, T_{\mathcal{SS}}^2(t_{*}, t).
\end{aligned}
\end{equation}
It is useful to note that the transfer functions themselves do not contain explicit wavenumber dependence, as long as the mode is in the super-horizon limit. The wavenumber-dependence is recovered, if one takes $t_*$ to be the horizon-crossing time for each mode. This implicit scale-dependence of the transfer functions can be used to compute the spectral index $n_s$ by taking derivatives of $T_{\cal RS}$ and $T_{\cal SS}$ with respect to $t_*$, as discussed for example in Ref.~\cite{KMS}.

\subsection{Evolution during angular inflation}

Having laid down the formulation of the evolution of perturbations, we are ready to proceed with the computation of the dynamical variables that control the evolution of the perturbations, from the turn rate $\omega$ to the isocurvature growth parameter $\tilde \beta$. We will again focus on the regime of small $\alpha$ and large $R_{\rm m}$, where significant analytical progress can be made. The intuition gained from this regime will be checked numerically and extended to the region of $\alpha \sim 1$. 

\begin{figure}[t]
\centering
 \includegraphics[width=0.48\textwidth]{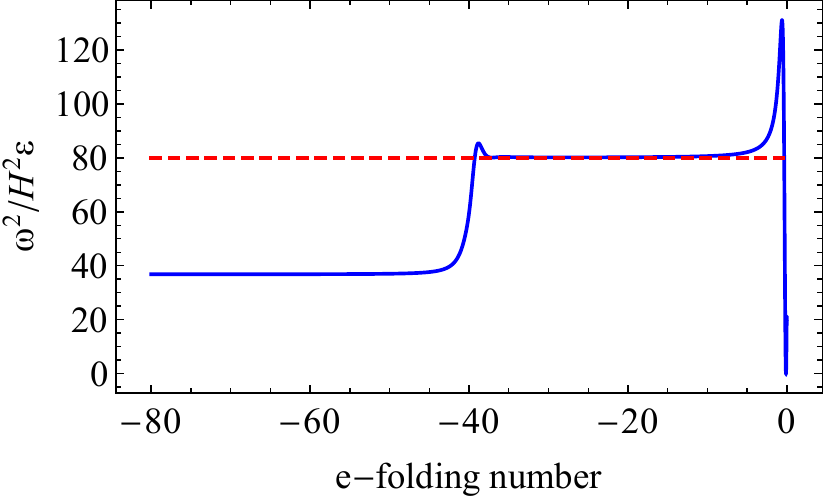} 
\includegraphics[width=0.48\textwidth]{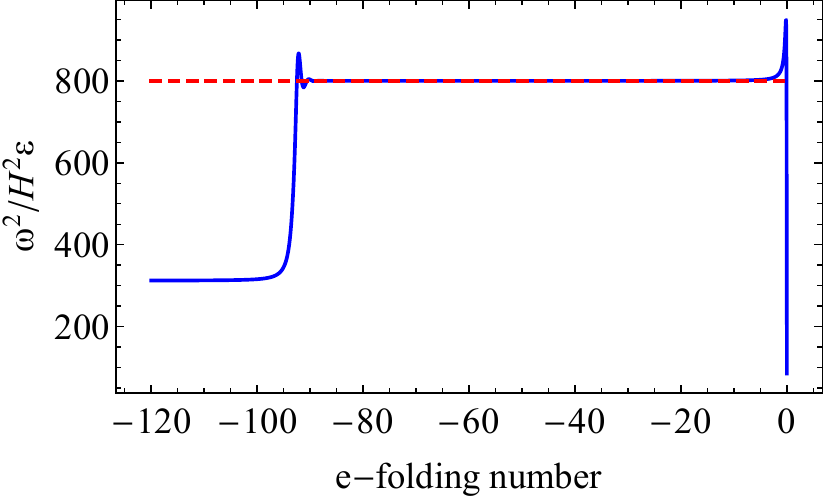} 
\caption{\it The normalized turn-rate $|\omega|/ (H\sqrt\epsilon)$, computed numerically in blue for $\alpha=0.016$, $R_{\rm m}=25$ (left) and $\alpha=0.0016$, $R_{\rm m}=9$ (right). The red-dashed lines correspond to $8/\alpha$, with excellent agreement in the angular regime (the last $40$ and $90$ $e$-folds respectively). Note that the same combination is constant  in the radial part (before the transition occurring halfway), as was found in Ref.~\cite{Achucarro:2017ing}.}
\label{fig:omegaoverHepsilon}
\end{figure}

First of all, using the fact that the motion occurs predominately along the angular direction, the tangent vector can be approximated by 
\begin{equation*}
\hat \sigma \approx \left( 0,\sqrt{ \cal G ^{\text{22}} }\text{sgn}(\dot \theta) \right) = \boldsymbol{\hat e_{\theta}} \, .
\end{equation*}
Hence the normalized turn rate vector will be along the radial direction, 
\begin{equation*}
\boldsymbol{\hat s} = \text{sgn}(\dot \theta) \boldsymbol{\hat e_r}= (\text{sgn}(\dot \theta) \sqrt{ \cal G ^{\text{11}}} ,0) \, .
\end{equation*}
It is easy to show that
\beqn
 {\omega^2={{\cal G}^{r r}V_r^2\over \dot\sigma^2}} \, .
\eeqn
Normalizing the turn-rate by the Hubble scale, we arrive at the relation
\beq
{{\omega^2 \over H^2 } = {4\over 3 \alpha} \epsilon} \, .
\eeq
We see that the turn rate is proportional to the curvature of the field-space manifold. It is illustrated in Figure \ref{fig:omegaoverHepsilon} that our analytic result agrees very well with the numerical evolution of the background fields. Moreover, using the expression for $\epsilon$ in the small $\alpha$ regime, we get
\beq
\omega^2 \simeq  {1 \over 3}(1-r^2) (\cos^2\theta + R_{\rm m} \sin^2\theta  )  \, .
\label{eq:omegaapprox}
\eeq  
The effective mass of the adiabatic perturbations during the angular evolution is
\beq
M_{\sigma\sigma} = \hat\sigma_I \hat\sigma^J {\cal M}^I_{~J}\simeq \hat\sigma^\theta\hat\sigma^\theta ({\cal D}_\theta{\cal D}_\theta V) \, .
\eeq
The adiabatic second-slow-roll parameter becomes
\beq
\eta_{\sigma\sigma} = {M_{\sigma\sigma}\over V }\simeq{ 2(1-r^2)\over 3 \alpha}\, .
\label{eq:msigmasigmaApprox1}
\eeq
By using Eq.~\eqref{eq:rthetaapprox2} this becomes
\beq
\eta_{\sigma\sigma} \simeq \frac{3 (\cot \theta + R_{\rm m} \tan \theta)^2}{(R_{\rm m}-1)^2}\, .
\label{eq:msigmasigmaApprox2}
\eeq  
Finally, using similar approximations we get
\beq
{\cal M}_{ss} = {\cal G}^{rr} ({\cal D}_r{\cal D}_r V) + \epsilon H^2 {\cal{ R}} \approx 
-{2 \over 3} (1-r^2) (\cos^2\theta + R_{\rm m} \sin^2\theta  )\, ,
\eeq 
where we took the isocurvature direction to be along the radial direction. We see that ${\cal M}_{ss}<0$, hence there is the possibility of the isocurvature modes being unstable. The corresponding slow-roll parameter is
\beq
\eta_{ss} = { M_{ss} \over V}  =-{4 ( 1-r^2)\over 3 \alpha}\, .
\label{eq:etassapprox}
\eeq
However, a closer look at Eq.~\eqref{ent} reveals that a large turn rate can still dominate the isocurvature effective mass. This is indeed true in our case, as we can see by comparing Eq.~\eqref{eq:etassapprox} to Eq~\eqref{eq:omegaapprox}, hence $\mu_s^2 >0$ and the isocurvature modes do not grow on super-horizon scales during  angular inflation.

\begin{figure}[t]
\centering
\includegraphics[width=0.47\textwidth]{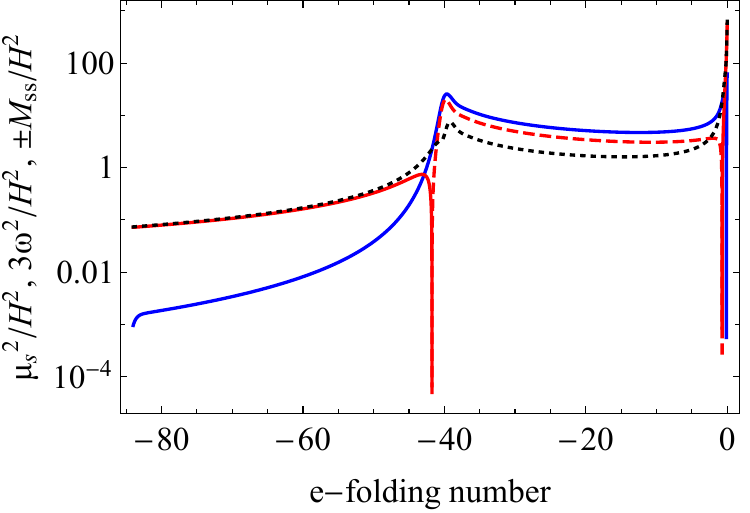}
\includegraphics[width=0.47\textwidth]{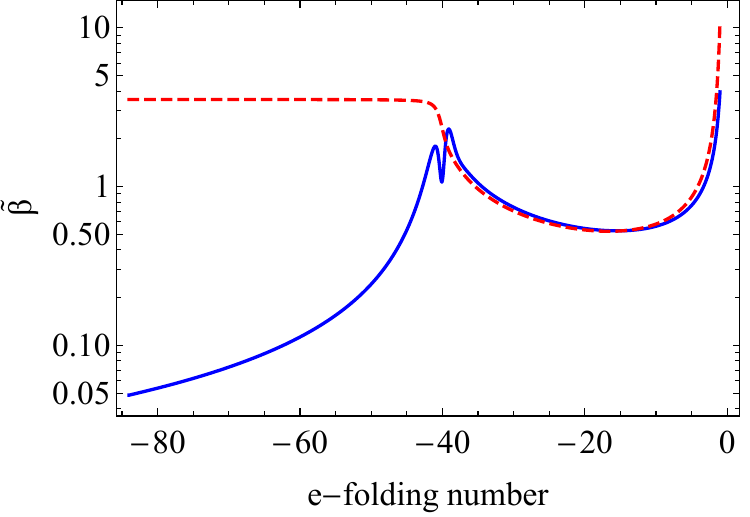}
\caption{\it 
(Left) 
The effective isocurvature mass-squared $\mu_s^2/H^2$ (black-dotted), along with the turn rate contribution $3\omega^2/H^2$ (blue) and the mass-squared ${\cal M}_{ss}/H^2$ (red) and $-{\cal M}_{ss}/H^2$ (red-dashed). All curves are computed numerically for $R_{\rm m}=25$ and $\alpha=0.016$, without using any approximations. We see that during angular inflation, which lasts for the final $40$ $e$-folds, $\mu_s^2/H^2 \sim O(1) >0$, while it is small and positive during the initial quasi-radial evolution.
(Right)
The isocurvature growth parameter $\tilde \beta$ for the same parameters numerically (blue-solid) and using the approximations of 
Eq.~\eqref{eq:betaapprox} (red-dashed). We see excellent agreement in the angular inflation regime.
}
\label{fig:betaplot}
\end{figure}

We now have all the components that go into the parameter $\tilde \beta$, which controls the growth ($\tilde\beta>0$) or decay ($\tilde\beta<0$) of the isocurvature modes, resulting in
\beq
\tilde\beta  \approx -\frac{3 (\cot \theta + R_{\rm m} \tan \theta)^2}{(R_{\rm m} - 1)^2} \,.
\label{eq:betaapprox}
\eeq   
This shows that the isocurvature modes will decay exponentially during the angular part of the inflationary trajectory. As seen in Fig.~\ref{fig:betaplot}, Eq.~\eqref{eq:betaapprox} is an excellent approximation to the numerical evaluation of $\tilde\beta$ in the angular inflation regime and in fact $|\tilde\beta| ={\cal O}(1)$, hinting at a severe suppression of isocurvature modes on super-horizon scales.
 Without an amplification, or at least a moderate decay, of the isocurvature modes, even a large turn-rate is insufficient to source any super-horizon evolution of the adiabatic modes. Hence, all adiabatic perturbation modes that have crossed the horizon before the onset of the period of angular inflation will be frozen and so will their spectral index $n_s$ and their amplitude that defines $r$.

\subsection{Superhorizon evolution and observables}\label{subsec:observables}

\begin{figure}[t]
\centering
\includegraphics[width=0.45\textwidth]{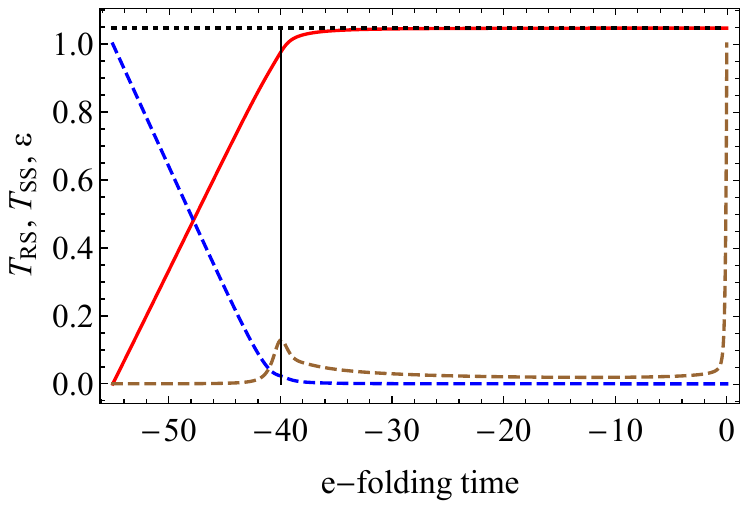}
\includegraphics[width=0.5\textwidth]{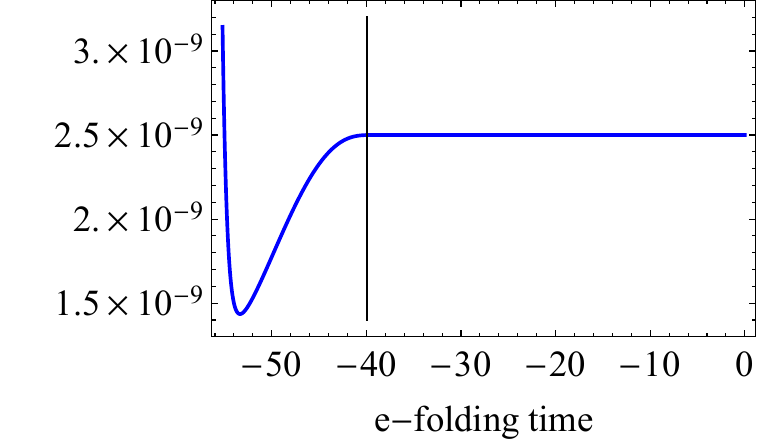}
\caption{ \it (Left) The transfer functions $T_{RS}$ (red) and $T_{SS}$ (dashed-blue) along with $\epsilon$ (dashed-brown) 
for $\alpha=0.016$ and $R_{\rm m}=25$.  The vertical line at $N\simeq -40$ signals the onset of the angular inflation period. We see that the transfer of power from the isocurvature to the adiabatic modes is negligible during angular inflation.
(Right) Evolution of the power spectrum for the same parameters.  The mode shown leaves the horizon $55$ $e$-folds before the end of inflation and the vertical line shows the onset of the angular inflation regime at $N\simeq -40$.
}
\label{fig:trsplot}
\end{figure}

With all the ingredients in place, it is a simple exercise to compute the super-horizon evolution of the adiabatic and isocurvature modes during the non-radial part of inflation. Since, as we showed, during angular inflation, $\tilde \beta <0$, we expect that the isocurvature modes will be quickly damped and hence the adiabatic modes will not be sourced. This is exactly what is shown in Fig.~\ref{fig:trsplot}, which presents a characteristic example from a larger number of simulated inflationary trajectories. We see a mild decay of the isocurvature modes during radial inflation, which can -through the non-zero turn rate- lead to a sourcing of the adiabatic modes \cite{Achucarro:2017ing}. However, once radial inflation ends and the system transitions into the angular regime, the isocurvature modes quickly decay and the transfer of power to the adiabatic modes ceases ($T_{\mathcal{RS}} = {\rm const}$). Looking at $\Delta T_{\mathcal{RS}}\equiv T_{\mathcal{RS}}(t_{\rm end}) - T_{\mathcal{RS}}$ we can see that $T_{\mathcal{ RS}}$ grows during the radial part, hence adiabatic modes are continuously sourced by isocurvature ones.
However, this sourcing stops immediately after the radial part has ended (hence $T_{\mathcal{RS}}$ stops increasing), signaling the fact that the curvature fluctuation has reached its adiabatic limit.
By taking the fiducial scale $t_*$ in Eq.~\eqref{eq:trans} close to the end of the initial period of radial inflation, $T_{\mathcal{SS}}$ is quickly forced towards zero and so $T_{\mathcal{RS}}$ becomes constant.

\begin{figure}[t]
\centering
\includegraphics[width=0.45\textwidth]{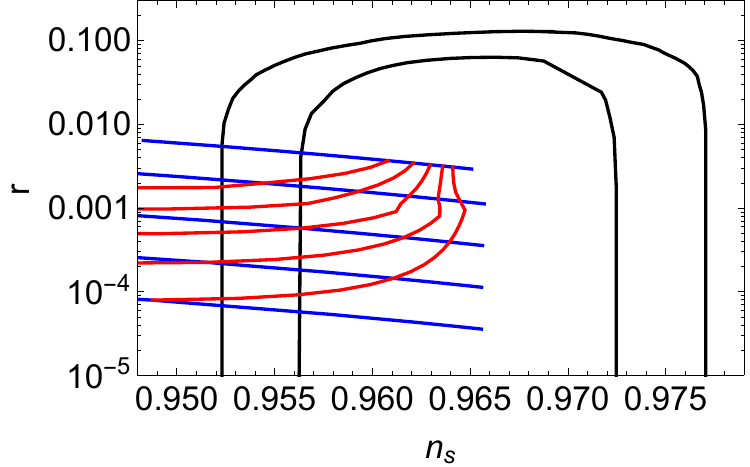}
\includegraphics[width=0.45\textwidth]{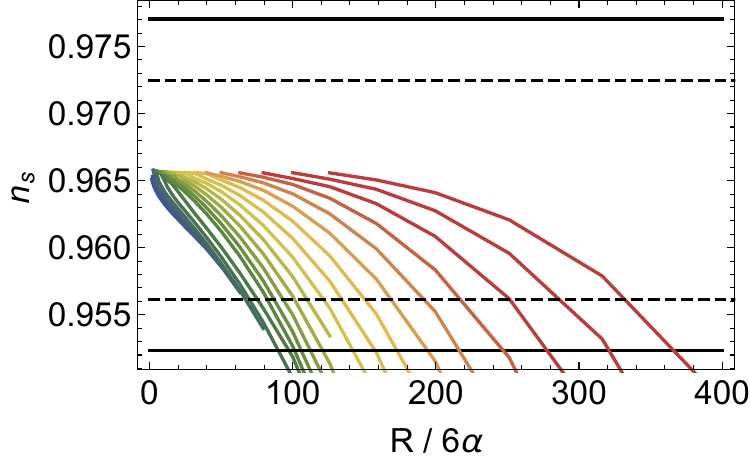}
\caption{\it 
(Left) The evolution of the primordial observables on the $n_s-r$ plane. 
Each blue line corresponds to a fixed value of $\alpha$ and varying $R_{\rm m}$ and is of the form $r = 3\alpha (n_s-1)^2$. The parameters are (from top to bottom) $6\alpha\simeq 0.8,0.3, 0.1,0.03,0.01$.
Each red line corresponds to a fixed value of $R=m_\chi^2 / m_\phi^2$ and varying $\alpha$.
The parameters are (from top to bottom) $R_{\rm m} \simeq 25, 16, 10, 6, 4$.
(Right) The scalar spectral index $n_s$ as a function of $R_{\rm m}/(6\alpha)$ for various values of $\alpha$, ranging from $6\alpha=0.01$ (blue) to $6\alpha\simeq 0.8$ (red) . The horizontal black-dashed and solid-black lines correspond to the Planck $1\sigma$ and $2\sigma$ intervals respectively. In both panels we took the CMB-relevant modes to have crossed the horizon $60$ $e$-folds before the end of inflation.
}
\label{fig:planck_Scan}
\end{figure}

In specific regimes of parameter space, the period of angular inflation can serve as a second phase of inflation that shifts the effective number of $e$-folds during the initial radial phase. This is in particular the case when the number of $e$-folds during angular inflation amounts to at most a few decades, such that the observable CMB window still takes place during the radial phase. This offset in e-folds due to the angular phase is given by $N(\theta_0)$ in Eq.~\eqref{eq:Nangular}, where the angle should be taken at the point where the inflationary trajectory joins the angular phase (which depends on the initial conditions). Due to the decay of the isocurvature modes and the lack of significant feeding into the adiabatic modes, the CMB predictions are therefore identical to those during the radial phase modulo the offset\footnote{We have also verified numerically these relations using \textbf{m.transport} \cite{Dias:2015rca}.}:
 \begin{align}
  n_s = 1 - \frac{2}{N - N(\theta_0)} \,, \qquad r = \frac{12 \alpha}{\left (N-N(\theta_0)\right)^2} \,, \label{offset}
 \end{align}
where $N\sim 55-60$ and $N(\theta_0)$ depends on initial conditions as well as the model parameters $\alpha$ and $R_{\rm m}$.
 Fig.~\ref{fig:planck_Scan} shows the evolution of the observables on the $n_s-r$ plane for varying field space curvature and varying mass asymmetry between the two fields.
We can see, as expected from the analysis of Section \ref{section:background}, that smaller $\alpha$ requires less mass asymmetry to deviate significantly from the single-field observables. Furthermore, for low values of $\alpha$, the curves in the left panel of Fig.~\ref{fig:planck_Scan} become degenerate (green to blue curves), hence the spectral index depends only on the combination $R_{\rm m}/\alpha$, as is expected from the leading term in the series expansion of Eq.~\eqref{eq:seriesN}. Going back to the contour plots of Fig.~\ref{fig:Ncontourplot}, for pairs of $R_{\rm m}$ and $\alpha$ that are below the two red lines, the observables are within the $1\sigma$ and $2\sigma$ regions of {\it Planck} respectively.  We also expect the local non-Gaussianity to be similarly affected by the angular inflation regime, retaining its form $f_{NL}\propto N^{-1}$ given in Ref.~\cite{Achucarro:2017ing}, with the substitution $N \to N-N(\theta)$.

Going to larger curvatures and mass ratios will lead to a number of angular e-folds $N(\theta)$ that exceeds the observable ones, implying that the CMB horizon crossing point is not during the radial but rather during the angular phase. This corresponds to the part of parameter space that lies above the thick black line of Fig.~\ref{fig:Ncontourplot}.
The calculation of inflationary observables during this phase is non-trivial, since the value of the turn-rate $\omega/H={\cal O}(1)$ makes the adiabatic and isocurvature perturbations coupled at sub-horizon scales \cite{Cremonini:2010ua}. We leave this analysis for a separate publication.

\section{Summary and Discussion} \label{sec:summary}

We have investigated multi-field $\alpha$ attractors in different regions of parameter space. For moderate values of the field-space curvature and mass ratio, inflation proceeds with a small turn-rate, which leads to a continuous but mild transfer of power from the isocurvature to the adiabatic mode. This regime was recently explored in Ref.~\cite{Achucarro:2017ing}, where it was found that the final values of the cosmological observables $n_s$ and $r$ do not deviate from their single-field counterparts.

The situation alters drastically when one increases the field-space curvature or potential gradients. In this case, as shown in Fig.~\ref{fig:Ncontourplot}, the system can show a large number of e-folds along a non-radial direction. This can be a combination of angular inflation, where the two fields move along the angular direction close to the boundary of the Poincare disk, and a period of inflation along the direction of the light field. Specifically, we have identified the attractor solution of Eq.~\eqref{eq:rthetafull} corresponding to angular inflation, which differs from the usual slow-roll one and is fully independent of initial conditions. This leads to the following sequence of phases:
\begin{itemize}
\item Starting with zero velocities the fields will accelerate towards the origin following an almost radial motion. The duration of this stage in $e$-folds is given by the usual single-field $\alpha$-attractor formula of Eq.~\eqref{nefsr}.
\item After a quick transient, which is very close to geodesic motion, the system is set into angular motion close to the boundary of the Poincare disk. There is a very accurate analytical description of the dynamics in this regime for larger values of the mass ratio and the curvature.
\item If the mass ratio is very large, the system can inflate along the direction of the light field for a significant amount of $e$-folds. This regime is well described as single-field or quasi-single-field evolution. 
\end{itemize}
For parameter values that lead to at most a number of decades of angular e-folds in the last two phases, the predictions for the CMB observables are altered in a very simple and intuitive way based on the offset with the additional angular phase \eqref{offset}.
This follows from our analysis in Section~\ref{sec:perturbations}: during the angular part of the evolution, the  isocurvature modes are quickly damped on super-horizon scales, leading to the effective freeze-out of the adiabatic modes. This means that the adiabatic modes retain their amplitude since the end of the radial part of inflation. 

An interesting question regards the analysis of and predictions for inflationary observables for the range of parameter space with all sixty e-folds of angular inflation, which will be presented separately.
Similarly, given the higher-dimensional moduli spaces of string theory, it would be interesting to analyze an $\alpha$-attractor model with more fields and a certain distribution of masses. Interesting questions include the duration of the angular phase as a function of the number of fields. Moreover, whether there is an emergent simplicity in the many-field limit, as has been shown for the flat geometry case \cite{Easther:2013rva}, remains to be seen. 

It is also interesting to contrast this behavior to non-minimally coupled models, which together with $\alpha$-attractors are among the most studied and well motivated inflationary models involving a non-trivial field space metric. In that case,  it has been shown that for generic initial conditions and randomly chosen potential parameters (constrained to provide enough $e$-folds of infllation and the observed amplitude of density perturbations), the cosmological observables, such as $n_s$ and $r$ fall within the {\it Planck} allowed region, on top of the predictions of the Starobinsky model and the single field predictions of $\alpha$-attractors, given in Eq.~\eqref{universal}. The ``lumpy" potential in simple models of quartic non-minimally coupled inflation, leads to a quick transient period, in which the fields quickly relax into a potential ``valley" and inflate along it, leading to largely parameter-independent predictions for the CMB observables \cite{KS}. In order to go beyond this universal behavior, one needs to fine tune the potential parameters and also fine-tune the initial conditions \cite{SSK}. Hence, in simple non-minimally coupled models, deviations from Eq.~\eqref{universal} require extreme fine-tuning, or some other mechanism, like a softly broken symmetry accompanied by an early phase of inflation ending at a waterfall transition, placing the field exponentially close to a potential ``ridge".  

It is quite remarkable that non-minimally coupled models and $\alpha$-attractors (with moderate parameter values) both respect their single-field predictions, even for multiple inflaton fields, however they do so in completely different ways: in non-minimally coupled models the large angular gradients force the fields into a single-field trajectory, while for $\alpha$ attractors the ``stretching" of field space along the angular directions allows for a quasi-single field trajectory with a small turn rate. Whether there is a unifying perspective on these types of behaviour remains to be seen. Finally, it is worth noting, that the evolution of observables during (p)reheating in multi-field models of inflation requires special attention \cite{Frolov, LythRiotto, SebastienRP, Taruya, FinelliBrandenberger,Tsujikawa:2000ab,Tsujikawa:2002nf, ChambersRajantie, BondFrolov, Elliston:2011dr, Leung:2012ve, Bassett:1998wg, Bassett:1999mt,  Bassett:1999ta, Amin:2014eta}, especially in the absence of a strong single-field attractor \cite{DeCross:2015uza,DeCross:2016fdz,DeCross:2016cbs}.

\section*{Acknowledgements}
 
 It is a pleasure to thank Ana Ach\'{u}carro, Renata Kallosh, Andrei Linde, S\'{e}bastien Renaux-Petel, Jacopo Fumagalli, Dong-Gang Wang and Yvette Welling for helpful discussions. The authors gratefully acknowledge support from the Dutch Organisation for Scientific Research (NWO). 

\appendix

\section{Useful geometric quantities}
For completeness we give the various geometric quantities of the curved manifold under consideration. The components of the field-space metric in polar coordinates are 
\begin{subequations}
\beqn
{\cal G}_{rr} &=& {6 \alpha\over (1-r^2)^2} \,, \qquad
{\cal G}_{\theta\theta}={6 \alpha\, r^2 \over (1-r^2 )^2} \,.
\eeqn
\end{subequations}
The Christoffel symbols are
\begin{subequations}
\beqn
\Gamma^r_{rr} &=&  {2r\over 1-r^2} \,, \qquad 
\Gamma^r_{\theta\theta} =-{r(1+r^2)\over 1-r^2} \,, \qquad 
\Gamma^\theta_{\theta r} ={1+r^2 \over r(1-r^2)} \,.
\eeqn
\end{subequations}
Finally, the Riemann tensor reads
\begin{subequations}
\beqn
&& {\cal R}^r_{\theta r\theta} = {-4 r^2 \over (1-r^2)^2} \,, \quad
{\cal R}^r_{\theta\theta r} = {4r^2 \over (1-r^2)^2} \,, \quad
{\cal R}^\theta_{r r\theta} = {4\over (1-r^2)^2} \,, \quad
{\cal R}^\theta_{r \theta r} = {-4\over (1-r^2)^2} \,. \notag
\eeqn
\end{subequations}
with the Ricci scalar given by ${\cal R}=-{4 / 3\alpha}$. Thus the field-space has a constant negative curvature controlled by $\alpha$, and its
geodesics (in the Poincare disk coordinates) are circles that intersect the boundary at a right angle.


\begin{thebibliography}{999}

\bibitem{Ade:2015lrj} 
  P.~A.~R.~Ade {\it et al.} [Planck Collaboration],
  ``Planck 2015 results. XX. Constraints on inflation,''
  Astron.\ Astrophys.\  {\bf 594}, A20 (2016)
  [arXiv:1502.02114 [astro-ph.CO]]. 

\bibitem{Guth:1980zm}
  A.~H.~Guth,
  ``The Inflationary Universe: A Possible Solution to the Horizon and Flatness Problems,''
  Phys.\ Rev.\ D {\bf 23} (1981) 347.

\bibitem{Linde:1981mu} 
  A.~D.~Linde,
  ``A New Inflationary Universe Scenario: A Possible Solution of the Horizon, Flatness, Homogeneity, Isotropy and Primordial Monopole Problems,''
  Phys.\ Lett.\  {\bf 108B}, 389 (1982).

\bibitem{Ade:2015tva} 
P.~A.~R.~Ade {\it et al.} [BICEP2 and Planck Collaborations],
``Joint Analysis of BICEP2/$Keck  Array$ and $Planck$ Data,''
Phys.\ Rev.\ Lett.\  {\bf 114}, 101301 (2015)
[arXiv:1502.00612 [astro-ph.CO]].



\bibitem{Peterson:2011yt}
C.~M.~Peterson and M.~Tegmark,
``Testing multifield inflation: A geometric approach,''
Phys. Rev. D \textbf{87} (2013) no.10, 103507
[arXiv:1111.0927 [astro-ph.CO]].


\bibitem{Peterson:2010np}
C.~M.~Peterson and M.~Tegmark,
``Testing Two-Field Inflation,''
Phys. Rev. D \textbf{83} (2011), 023522
[arXiv:1005.4056 [astro-ph.CO]].

\bibitem{Peterson:2010mv}
C.~M.~Peterson and M.~Tegmark,
``Non-Gaussianity in Two-Field Inflation,''
Phys. Rev. D \textbf{84} (2011), 023520
[arXiv:1011.6675 [astro-ph.CO]].



\bibitem{heavy1}
  A.~Ach\'{u}carro, J.~O.~Gong, S.~Hardeman, G.~A.~Palma and S.~P.~Patil,
  ``Effective theories of single field inflation when heavy fields matter,''
  JHEP {\bf 1205} (2012) 066
  [arXiv:1201.6342 [hep-th]].

\bibitem{heavy2}
  A.~Ach\'{u}carro, V.~Atal, S.~Cespedes, J.~O.~Gong, G.~A.~Palma and S.~P.~Patil,
  ``Heavy fields, reduced speeds of sound and decoupling during inflation,''
  Phys.\ Rev.\ D {\bf 86} (2012) 121301
  [arXiv:1205.0710 [hep-th]].
  
  \bibitem{Achucarro:2016fby} 
  A.~Ach\'{u}carro, V.~Atal, C.~Germani and G.~A.~Palma,
``Cumulative effects in inflation with ultra-light entropy modes,''
  JCAP {\bf 1702}, no. 02, 013 (2017)
  [arXiv:1607.08609 [astro-ph.CO]].
  
  
  
  
\bibitem{Renaux-Petel:2015mga}
  S.~Renaux-Petel and K.~Turzyński,
  ``Geometrical Destabilization of Inflation,''
  Phys.\ Rev.\ Lett.\  {\bf 117} (2016) no.14,  141301
  [arXiv:1510.01281 [astro-ph.CO]].

\bibitem{Kallosh:2013yoa} 
  R.~Kallosh, A.~Linde and D.~Roest,
  ``Superconformal Inflationary $\alpha$-Attractors,''
  JHEP {\bf 1311}, 198 (2013)
  [arXiv:1311.0472 [hep-th]].
 
 
\bibitem{Ferrara:2013rsa}
S.~Ferrara, R.~Kallosh, A.~Linde and M.~Porrati,
``Minimal Supergravity Models of Inflation,''
Phys. Rev. D \textbf{88} (2013) no.8, 085038
[arXiv:1307.7696 [hep-th]].

\bibitem{Kallosh:2013maa}
R.~Kallosh and A.~Linde,
``Non-minimal Inflationary Attractors,''
JCAP \textbf{10} (2013), 033
[arXiv:1307.7938 [hep-th]].

\bibitem{Kallosh:2013daa}
R.~Kallosh and A.~Linde,
``Multi-field Conformal Cosmological Attractors,''
JCAP \textbf{12} (2013), 006
[arXiv:1309.2015 [hep-th]].

\bibitem{Kallosh:2015lwa}
R.~Kallosh and A.~Linde,
``Planck, LHC, and $\alpha$-attractors,''
Phys. Rev. D \textbf{91} (2015), 083528
[arXiv:1502.07733 [astro-ph.CO]].

\bibitem{Carrasco:2015rva}
J.~J.~M.~Carrasco, R.~Kallosh and A.~Linde,
``Cosmological Attractors and Initial Conditions for Inflation,''
Phys. Rev. D \textbf{92} (2015) no.6, 063519
[arXiv:1506.00936 [hep-th]].




\bibitem{Staro}
  A.~A.~Starobinsky,
  ``A New Type of Isotropic Cosmological Models Without Singularity,''
  Phys.\ Lett.\  {\bf 91B} (1980) 99.
  
  


\bibitem{nonm1} R. Kallosh, A. Linde, and D. Roest, ``Universal attractor for inflation at strong coupling," Phys. Rev. Lett. 112 (2014): 011303, arXiv:1310.3950 [hep-th].



\bibitem{KS} D. I. Kaiser and E. I. Sfakianakis, ``Multifield inflation after Planck: The case for nonminimal couplings," Phys. Rev. Lett. 112 (2014): 011302, arXiv:1304.0363 [astro-ph.CO].


\bibitem{Galante:2014ifa} 
  M.~Galante, R.~Kallosh, A.~Linde and D.~Roest,
  Phys.\ Rev.\ Lett.\  {\bf 114}, no. 14, 141302 (2015)
  [arXiv:1412.3797 [hep-th]].

\bibitem{Carrasco}
  J.~J.~M.~Carrasco, R.~Kallosh, A.~Linde and D.~Roest,
  ``Hyperbolic geometry of cosmological attractors,''
  Phys.\ Rev.\ D {\bf 92} (2015) no.4,  041301
  [arXiv:1504.05557 [hep-th]].



\bibitem{Dimopoulos:2005ac} 
  S.~Dimopoulos, S.~Kachru, J.~McGreevy and J.~G.~Wacker,
  ``N-flation,''
  JCAP {\bf 0808}, 003 (2008)
  [hep-th/0507205].

\bibitem{Easther:2013rva} 
  R.~Easther, J.~Frazer, H.~V.~Peiris and L.~C.~Price,
  ``Simple predictions from multifield inflationary models,''
  Phys.\ Rev.\ Lett.\  {\bf 112}, 161302 (2014)
  [arXiv:1312.4035 [astro-ph.CO]].

\bibitem{Dias:2016slx} 
  M.~Dias, J.~Frazer and M.~C.~D.~Marsh,
  ``Simple emergent power spectra from complex inflationary physics,''
  Phys.\ Rev.\ Lett.\  {\bf 117}, no. 14, 141303 (2016)
  [arXiv:1604.05970 [astro-ph.CO]].



\bibitem{Dias:2017gva} 
  M.~Dias, J.~Frazer and M.~c.~D.~Marsh,
  ``Seven Lessons from Manyfield Inflation in Random Potentials,''
  arXiv:1706.03774 [astro-ph.CO].

\bibitem{Easther:2005zr} 
  R.~Easther and L.~McAllister,
  ``Random matrices and the spectrum of N-flation,''
  JCAP {\bf 0605}, 018 (2006)
  [hep-th/0512102].


\bibitem{Achucarro:2017ing} 
  A.~Achucarro, R.~Kallosh, A.~Linde, D.~G.~Wang and Y.~Welling,
``Universality of multi-field $\alpha$-attractors,''
  arXiv:1711.09478 [hep-th].


\bibitem{Brown:2017osf} 
  A.~R.~Brown,
  ``Hyperinflation,''
  arXiv:1705.03023 [hep-th].
  
\bibitem{Mizuno:2017idt} 
  S.~Mizuno and S.~Mukohyama,
  ``Primordial perturbations from inflation with a hyperbolic field-space,''
  Phys.\ Rev.\ D {\bf 96}, no. 10, 103533 (2017)
  [arXiv:1707.05125 [hep-th]].





\bibitem{Linde:2018hmx}
A.~Linde, D.~G.~Wang, Y.~Welling, Y.~Yamada and A.~Achúcarro,
``Hypernatural inflation,''
JCAP {\bf 1807} (2018) no.07,  035
[arXiv:1803.09911 [hep-th]].





\bibitem{Frazer:2013zoa}
  J.~Frazer,
  ``Predictions in multifield models of inflation,''
  JCAP {\bf 1401} (2014) 028
  [arXiv:1303.3611 [astro-ph.CO]].


\bibitem{Lyth:2009zz} 
  D.~H.~Lyth and A.~R.~Liddle,
  ``The primordial density perturbation: Cosmology, inflation and the origin of structure,''
  Cambridge, UK: Cambridge Univ. Pr. (2009) 497 p


\bibitem{Yang:2012bs} 
  I.~S.~Yang,
 ``The Strong Multifield Slowroll Condition and Spiral Inflation,''
  Phys.\ Rev.\ D {\bf 85}, 123532 (2012)
  [arXiv:1202.3388 [hep-th]].


\bibitem{GarciaBellido:1995qq} 
  J.~Garcia-Bellido and D.~Wands,
 ``Metric perturbations in two field inflation,''
  Phys.\ Rev.\ D {\bf 53}, 5437 (1996)
  [astro-ph/9511029].

\bibitem{Starobinsky:1986fxa} 
  A.~A.~Starobinsky,
``Multicomponent de Sitter (Inflationary) Stages and the Generation of Perturbations,''
  JETP Lett.\  {\bf 42}, 152 (1985)
  [Pisma Zh.\ Eksp.\ Teor.\ Fiz.\  {\bf 42}, 124 (1985)].


\bibitem{Bond:2006nc} 
J.~R.~Bond, L.~Kofman, S.~Prokushkin and P.~M.~Vaudrevange,
``Roulette inflation with Kahler moduli and their axions,''
Phys.\ Rev.\ D {\bf 75}, 123511 (2007)
[hep-th/0612197].

\bibitem{Kallosh:2014qta} 
R.~Kallosh, A.~Linde, B.~Vercnocke and W.~Chemissany,
``Is Imaginary Starobinsky Model Real?,''
JCAP {\bf 1407}, 053 (2014)
[arXiv:1403.7189 [hep-th]].


\bibitem{Lazaroiu:2017avq} 
C.~I.~Lazaroiu and C.~S.~Shahbazi,
``Generalized two-field $\alpha$-attractor models from geometrically finite hyperbolic surfaces,''
Nucl.\ Phys.\ B {\bf 936}, 542 (2018)
doi:10.1016/j.nuclphysb.2018.09.018
[arXiv:1702.06484 [hep-th]].

\bibitem{KMS} D. I. Kaiser, E. A. Mazenc, and E. I. Sfakianakis, ``Primordial bispectrum from multifield inflation with nonminimal couplings," Phys. Rev. D87 (2013): 064004, arXiv:1210.7487 [astro-ph.CO].

\bibitem{GKS} R. N. Greenwood, D. I. Kaiser, and E. I. Sfakianakis, ``Multifield dynamics of Higgs inflation," Phys. Rev. D87 (2013): 044038, arXiv:1210.8190 [hep-ph].

\bibitem{SSK} K. Schutz, E. I. Sfakianakis, and D. I. Kaiser, ``Multifield inflation after Planck: Isocurvature modes from nonminimal couplings," Phys. Rev. D89 (2014): 064044, arXiv:1310.8285 [astro-ph.CO].

\bibitem{GongMultifield} J.-O. Gong, ``Multi-field inflation and cosmological perturbations," arXiv:1606.06971 [astro-ph.CO].


\bibitem{Bassett:2005xm} 
  B.~A.~Bassett, S.~Tsujikawa and D.~Wands,
``Inflation dynamics and reheating,''
  Rev.\ Mod.\ Phys.\  {\bf 78}, 537 (2006)
  [astro-ph/0507632].



\bibitem{Malik:2008im} 
  K.~A.~Malik and D.~Wands,
``Cosmological perturbations,''
  Phys.\ Rept.\  {\bf 475}, 1 (2009)
  [arXiv:0809.4944 [astro-ph]].


\bibitem{Gordon:2000hv}
C.~Gordon, D.~Wands, B.~A.~Bassett and R.~Maartens,
``Adiabatic and entropy perturbations from inflation,''
Phys. Rev. D \textbf{63}, 023506 (2000)
[arXiv:astro-ph/0009131 [astro-ph]].


\bibitem{GrootNibbelink:2000vx}
S.~Groot Nibbelink and B.~J.~W.~van Tent,
``Density perturbations arising from multiple field slow roll inflation,''
[arXiv:hep-ph/0011325 [hep-ph]].

\bibitem{GrootNibbelink:2001qt}
S.~Groot Nibbelink and B.~J.~W.~van Tent,
``Scalar perturbations during multiple field slow-roll inflation,''
Class. Quant. Grav. \textbf{19}, 613-640 (2002)
[arXiv:hep-ph/0107272 [hep-ph]].

  
  \bibitem{Wands:2007bd} 
  D.~Wands,
  ``Multiple field inflation,''
  Lect.\ Notes Phys.\  {\bf 738}, 275 (2008)
  [astro-ph/0702187 [ASTRO-PH]].
  
  \bibitem{Wands:2002bn} 
  D.~Wands, N.~Bartolo, S.~Matarrese and A.~Riotto,
  ``An Observational test of two-field inflation,''
  Phys.\ Rev.\ D {\bf 66}, 043520 (2002)
  [astro-ph/0205253].
  
  
\bibitem{KaiserTodhunter} D. I. Kaiser and A. T. Todhunter, ``Primordial perturbations from multifield inflation with nonminimal couplings," Phys. Rev. D 81 (2010): 124037, arXiv:1004.3805 [astro-ph.CO].


\bibitem{Dias:2015rca} 
M.~Dias, J.~Frazer and D.~Seery,
``Computing observables in curved multifield models of inflation—A guide (with code) to the transport method,''
JCAP {\bf 1512}, no. 12, 030 (2015)
[arXiv:1502.03125 [astro-ph.CO]].




\bibitem{Cremonini:2010ua} 
  S.~Cremonini, Z.~Lalak and K.~Turzynski,
  ``Strongly Coupled Perturbations in Two-Field Inflationary Models,''
  JCAP {\bf 1103}, 016 (2011)
  [arXiv:1010.3021 [hep-th]].


\bibitem{Frolov} A. V. Frolov, ``Non-linear dynamics and primordial curvature perturbations from preheating," Class. Quant. Grav. 27 (2010): 124006, arXiv:1004.3559 [gr-qc].


\bibitem{LythRiotto} D. H. Lyth and A. Riotto, ``Particle physics models of inflation and the cosmological density perturbation," Phys. Rept. 314 (1999): 1, arXiv:hep-ph/9807278.

\bibitem{SebastienRP}  S.~Renaux-Petel and K.~Turzy\'{n}ski, ``On reaching the adiabatic limit in multi-field inflation," arXiv:1405.6195 [astro-ph.CO].



\bibitem{Taruya} A. Taruya and Y. Nambu, ``Cosmological perturbation with two scalar fields in reheating after inflation," Phys. Lett. B 428 (1998): 37, arXiv:gr-qc/9709035.


\bibitem{FinelliBrandenberger} F. Finelli and R. Brandenberger, ``Parametric amplification of metric fluctuations during reheating in two field models," Phys. Rev. D 62 (2000): 083502, arXiv:hep-ph/0003172.



\bibitem{Tsujikawa:2000ab}
S.~Tsujikawa and B.~A.~Bassett,
``A New twist to preheating,''
Phys. Rev. D \textbf{62}, 043510 (2000)
[arXiv:hep-ph/0003068 [hep-ph]].


\bibitem{Tsujikawa:2002nf}
S.~Tsujikawa and B.~A.~Bassett,
``When can preheating affect the CMB?,''
Phys. Lett. B \textbf{536}, 9-17 (2002)
[arXiv:astro-ph/0204031 [astro-ph]].


\bibitem{ChambersRajantie} A. Chambers and A. Rajantie, ``Lattice calculation of non-Gaussianity from preheating," Phys. Rev. Lett. 100 (2008): 041302, arXiv:0710.4133 [astro-ph].

\bibitem{BondFrolov} J. R. Bond, A. V. Frolov, Z. Huang, and L. Kofman, ``Non-Gaussian spikes from chaotic billiards in inflation preheating," Phys. Rev. Lett. 103 (2009): 071301, arXiv:0903.3407 [astro-ph].




\bibitem{Elliston:2011dr}  J.~Elliston, D.~J.~Mulryne, D.~Seery and R.~Tavakol, ``Evolution of $f_{NL}$ to the adiabatic limit,'' JCAP 1111 (2011): 005, arXiv:1106.2153 [astro-ph.CO].

\bibitem{Leung:2012ve}  G.~Leung, E.~R.~M.~Tarrant, C.~T.~Byrnes and E.~J.~Copeland, ``Reheating, multifield inflation and the fate of the primordial observables,'' JCAP 1209 (2012): 008, arXiv:1206.5196 [astro-ph.CO].

\bibitem{Bassett:1998wg} 
  B.~A.~Bassett, D.~I.~Kaiser and R.~Maartens,
``General relativistic preheating after inflation,''
  Phys.\ Lett.\ B {\bf 455}, 84 (1999)
  [hep-ph/9808404].
  
  
  \bibitem{Bassett:1999mt} 
  B.~A.~Bassett, F.~Tamburini, D.~I.~Kaiser and R.~Maartens,
``Metric preheating and limitations of linearized gravity. 2.,''
  Nucl.\ Phys.\ B {\bf 561}, 188 (1999)
  [hep-ph/9901319].
  
  \bibitem{Bassett:1999ta} 
  B.~A.~Bassett, C.~Gordon, R.~Maartens and D.~I.~Kaiser,
``Restoring the sting to metric preheating,''
  Phys.\ Rev.\ D {\bf 61}, 061302 (2000)
  [hep-ph/9909482].
  
  
\bibitem{Amin:2014eta} 
  M.~A.~Amin, M.~P.~Hertzberg, D.~I.~Kaiser and J.~Karouby,
``Nonperturbative Dynamics Of Reheating After Inflation: A Review,''
  Int.\ J.\ Mod.\ Phys.\ D {\bf 24}, 1530003 (2014)
  [arXiv:1410.3808 [hep-ph]].




\bibitem{DeCross:2015uza}
M.~P.~DeCross, D.~I.~Kaiser, A.~Prabhu, C.~Prescod-Weinstein and E.~I.~Sfakianakis,
``Preheating after Multifield Inflation with Nonminimal Couplings, I: Covariant Formalism and Attractor Behavior,''
Phys. Rev. D \textbf{97}, no.2, 023526 (2018)
[arXiv:1510.08553 [astro-ph.CO]].

  
\bibitem{DeCross:2016fdz}
M.~P.~DeCross, D.~I.~Kaiser, A.~Prabhu, C.~Prescod-Weinstein and E.~I.~Sfakianakis,
``Preheating after multifield inflation with nonminimal couplings, II: Resonance Structure,''
Phys. Rev. D \textbf{97}, no.2, 023527 (2018)
[arXiv:1610.08868 [astro-ph.CO]].
  

\bibitem{DeCross:2016cbs}
M.~P.~DeCross, D.~I.~Kaiser, A.~Prabhu, C.~Prescod-Weinstein and E.~I.~Sfakianakis,
``Preheating after multifield inflation with nonminimal couplings, III: Dynamical spacetime results,''
Phys. Rev. D \textbf{97}, no.2, 023528 (2018)
[arXiv:1610.08916 [astro-ph.CO]].

  
\end{thebibliography}
\end{document}